\documentclass{article} 

\usepackage{bm}
\usepackage{mathtools}
\usepackage{amsmath}
\usepackage{amssymb}
\usepackage[a4paper, total={6in, 8in}]{geometry}

\newcommand{\Nu}{{\rm Nu}}   
\newcommand{\Rey}{{\rm Re}}   
\newcommand{\Ra}{{\rm Ra}}   
\newcommand{\Pra}{{\rm Pr}}   
\newcommand{\ur}{\tilde{u}_{\rm rms}}

\title{\textbf{Connecting boundary layer dynamics with extreme bulk dissipation events in Rayleigh-B\'{e}nard flow}}

\author{Valentina Valori$^{1}$ and J\"org Schumacher$^{1,2}$}

\date{%
    $^1$Institut f\"ur Thermo- und Fluiddynamik, Technische Universit\"at Ilmenau, D-98684 Ilmenau, Germany\\%
    $^2$Tandon School of Engineering, New York University, New York City, NY 11201, USA\\[2ex]%
    \today
}

\begin{document}

\maketitle

\section{Abstract}
We study the connection between extreme events of thermal and kinetic energy dissipation rates in the bulk of three-dimensional Rayleigh-B\'{e}nard convection and the wall shear stress patterns at the top and the bottom planes that enclose the layer. Zero points of this two-dimensional vector field stand for detachments of strong thermal plumes. If their position at the opposite planes and a given time is close then they can be considered as precursors for high-amplitude bulk dissipation events triggered by plume collisions or close passings. This scenario requires a breaking of the synchronicity of the boundary layer dynamics at both plates which is found to be in line with a transition of the bulk derivative statistics from Gaussian to intermittent. Our studies are based on three-dimensional high-resolution direct numerical simulations for moderate Rayleigh numbers between $\Ra=10^4$ and $5\times 10^5$.\\

\section{Introduction} The relation between the dynamics in the boundary layers of wall-bounded flows and the statistical properties of the turbulence fields and their derivatives in the bulk is essential for a better understanding of turbulent transport and its modeling in numerous applications \cite{Prandtl1925,Spiegel1963,Howard1972}. Turbulent shear flows at large Reynolds numbers for example are known to generate near-wall streaks and vortices due to the non-normal transient amplification of fluctuations \cite{Adrian2007}. These near-wall structures are thought to be the nuclei and building blocks of larger-scale features in the flow further away from the wall such as large-scale motions or very-large scale motions \cite{Marusic2010,Smits2011,Feldmann2018}. Near-wall structure formation processes have been also been brought in connection with the topology of the wall shear stress vector field in the past years, in particular with zero points of this field \cite{Chong2012,Lenaers2012} which can generate local flow reversals.

In thermal convection, in its simplest setting a turbulent flow in a layer which is driven by buoyancy differences enclosed between two impermeable isothermal parallel plates, the most important near-wall structure is the thermal plume -- a detached fragment of the thermal boundary layer that rises or sinks into the bulk of the layer and undergoes (partial) turbulent mixing. Thermal plumes drive a whole cascade of convective motion and generate fluctuations of the turbulent fields and their derivatives in the bulk, particularly on smaller scales as the mean thickness of the thermal boundary layer $\delta_T$ \cite{Lohse2010,Verma2018}. The magnitudes of these fluctuations are essential in many applications as they determine turbulent transport coefficients, such as turbulent eddy viscosities and diffusivities. More detailed, a local eddy or subgrid-scale viscosity would require information on the velocity fluctuations $u_{\rm rms}=\langle u_i^2\rangle^{1/2}$ and the energy dissipation rate $\varepsilon$ to get $\nu_t (\bm{x},t)\sim u_{\rm rms}^2/\varepsilon$ \cite{Smagorisnky1958,Wilcox2006}.   One prominent example is solar convection where the properties of the bulk turbulence below the local scale heights are still unexplored but important for the transport and the generation of large-scale magnetic fields \cite{Kupka2017,Schumacher2020}.                        

In this letter, we want to investigate the link between derivative statistics in the bulk and the formation of plumes and plume clusters in the boundary layers in a Rayleigh-B\'{e}nard layer of height $H$ and size $8H\times 8H\times H$. Our focus is on far tail events of the probability density functions of the rates at which thermal variance and kinetic energy are dissipated, $\varepsilon_T$ and $\varepsilon$. The study is based on a record of very high-resolved direct numerical simulations that apply an exponentially fast converging spectral element technique. The range of Rayleigh numbers is moderate with values $1.5\times 10^4 \le \Ra \le 5\times 10^5$ such that moments of velocity and temperature derivatives up to order 8 can be resolved, which corresponds to an order of 4 of moments of the dissipation rates. We demonstrate that such high-amplitude dissipation events (which are the extreme events of the convective turbulence in the bulk) are caused by collisions or close passings of rising and falling thermal plumes in the center of the layer. This process becomes increasingly probable once the Rayleigh number is sufficiently large such that plume impact at one wall and plume detachment at the opposite wall get disconnected of each other. We term this the breaking of the boundary layer synchronicity. The indicator (or precursor) for such an event is the two-dimensional vector field $\bm{\tau}_w$ which is spanned by the wall shear stress components. Bandaru et al. \cite{Bandaru2015} considered $\bm{\tau}_w$ as a blue print of the boundary layer dynamics and saddle-type zero points are found to coincide with regions where the strongest plumes detach (see also \cite{Schumacher2016a,Pandey2018}).     

\section{Direct numerical simulations}
The three-dimensional Navier-Stokes and temperature equations for Rayleigh-B\'{e}nard convection were solved in the Boussinesq approximation in direct numerical simulations (DNS) in a Cartesian cell with horizontal coordinates $x$ and $y$ and vertical coordinate $z$ (parallel to the acceleration due to gravity ${\bm g}=(0,0,-g)$) with an aspect ratio $\Gamma=L/H=8$. The equations for the velocity field ${\bm u}({\bm x},t)$ and the temperature field $T({\bm x},t)$ are given by
\begin{align}
\nabla\cdot \bm{u} &= 0,\\
\frac{\partial \bm{u}}{\partial t} + (\bm{u}\cdot\nabla)\bm{u} &= -\frac{1}{\rho_{0}}\nabla p + \nu\nabla^{2}\bm{u} + g\alpha(T-T_{0})\bm{e}_{z},\\
\frac{\partial T}{\partial t} + (\bm{u}\cdot\nabla)T &= \kappa\nabla^{2}T.
\end{align}
Here, $\nu$ is the kinematic viscosity, $\kappa$ the thermal diffusivity, $\rho_0$ the constant reference mass density, $T_0$ the reference temperature, $p$ the pressure deviation, and $\alpha$ the thermal expansion coefficient.  In dimensionless form, all length scales are expressed in units of height $H$, all velocities in units of the free-fall velocity $U_f = \sqrt{g\alpha\Delta T H}$, and all temperatures in units of $\Delta T=T_{\rm bottom}-T_{\rm top}$. Periodic boundary conditions were applied at all side walls for both, the velocity and the temperature field. At the horizontal walls no-slip boundary conditions were applied for the velocity field and isothermal boundary conditions for the temperature field. The Prandtl number is $\Pra=\nu/\kappa=1$ and the Rayleigh number $\Ra=g\alpha \Delta T H^3/(\nu\kappa)$. The Rayleigh number varies between $1.5\times 10^4$ and $5\times 10^5$ for most of the analysis. We have further DNS data for our $\Ra\ge 1900$ which cover the weakly nonlinear regime at the onset of convection. For these very low Rayleigh numbers the flow is steady and develops small-amplitude fluctuations only.
\begin{figure}
\begin{center}
\includegraphics[width=0.6\textwidth]{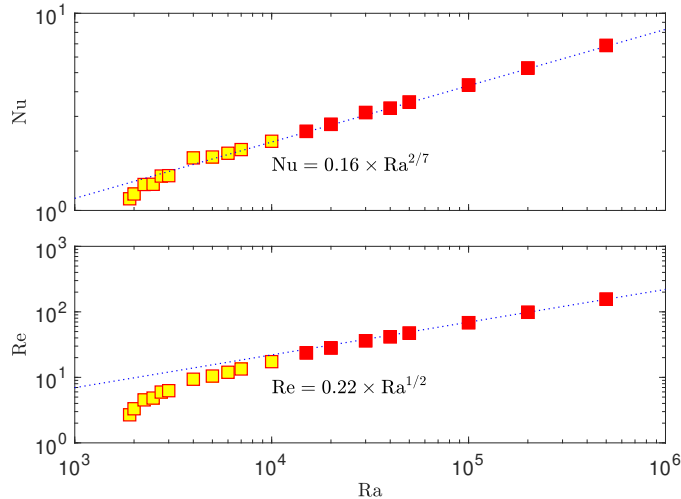}
\caption{Global transport properties of the turbulent convection flow. Top: Nusselt number. Bottom: Reynolds number. Data are for Rayleigh numbers $1.9\times 10^3\le Ra\le 5\times 10^5$. The filled data points for $1.5\times 10^4\le Ra\le 5\times 10^5$ will be considered for the subsequent analysis. The power law fits to the data are indicated by the dotted lines.}
\label{transp}
\end{center}
\end{figure}
\begin{figure}
\begin{center}
\includegraphics[width=0.6\textwidth]{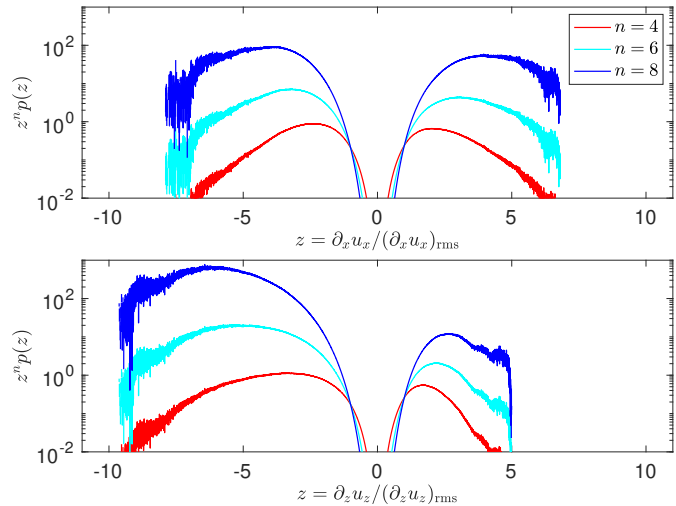}
\caption{Statistical convergence of longitudinal derivatives at $\Ra=5\times 10^5$. The area under the curve $z^n p(z)$ corresponds to the normalized $n$th-order derivative moment $M_n$ given by \eqref{Mnf}. Orders are listed in the legend in the top panel. The definition of the dimensionless variable $z$ is given in the axis label.
\label{conv}}
\end{center}
\end{figure}

The global transport of heat and momentum  displayed in Fig. \ref{transp}. These are the dimensionless Nusselt number $\Nu=1+\sqrt{\Ra} \langle \tilde{u}_z\tilde{T}\rangle_{V,t}$ which measures heat transfer and the Reynolds number $\Rey=\sqrt{\Ra}\,\ur$ which measures momentum transfer and contains the dimensionless root mean square velocity $\ur=\langle\tilde{u}_x^2+\tilde{u}_y^2+\tilde{u}_z^2\rangle^{1/2}_{V,t}$. Here, $\langle \cdot\rangle_{V,t}$ is a combined volume and time average. The turbulent heat transfer follows an algebraic power law that fits to a very good approximation a 2/7 law for the present range of Rayleigh numbers \cite{Johnston2009,Castaing1989}. 

The equations are solved by a spectral element method using the Nek5000 software package \cite{Fischer1997,Scheel2013}. The domain is covered by 819,200 spectral elements with a polynomial order of $N=5$ for the lower Rayleigh numbers up to $N=9$ for the largest Rayleigh number. These resolutions are required to obtain converged high-order derivative statistics. Beside the larger aspect ratio this is the limiting factor for the accessible Rayleigh number. Each individual simulation run started from the diffusive equilibrium plus small-amplitude white-in-time noise. Statistical properties are monitored once the flow is relaxed into a statistically stationary regime. The statistical analysis is done for most cases over 60 equidistant full snapshots. For the highest Rayleigh number they are separated by $2.9 T_f=2.9 H/U_f$ of each other. For two cases a sequence of snapshots was written out at higher frequency to study the formation of high-amplitude dissipation events. Figure \ref{conv} demonstrates the statistical convergence of derivative moments of orders $n=4$, 6 and 8 for two longitudinal derivatives at the highest Rayleigh number $\Ra=5\times 10^5$. We therefore plot $z^n p(z)$ versus $z$ with $z=\partial_{\tilde{x}} \tilde{u}_x/(\partial_{\tilde{x}} \tilde{u}_x)_{\rm rms}$ (top) and $z=\partial_{\tilde{z}} \tilde{u}_z/(\partial_{\tilde{z}} \tilde{u}_z)_{\rm rms}$. The abbreviation rms stands for the root mean square value and $p(z)$ is the probability density function (PDF) of the normalized quantity $z$. For the remaining text we omit the tilde for dimensionless quantities to ease the notation. It is stated when quantities with a physical dimension are used.
 
\section{Derivative moments} 
\begin{figure}
\begin{center}
\includegraphics[width=0.7\textwidth]{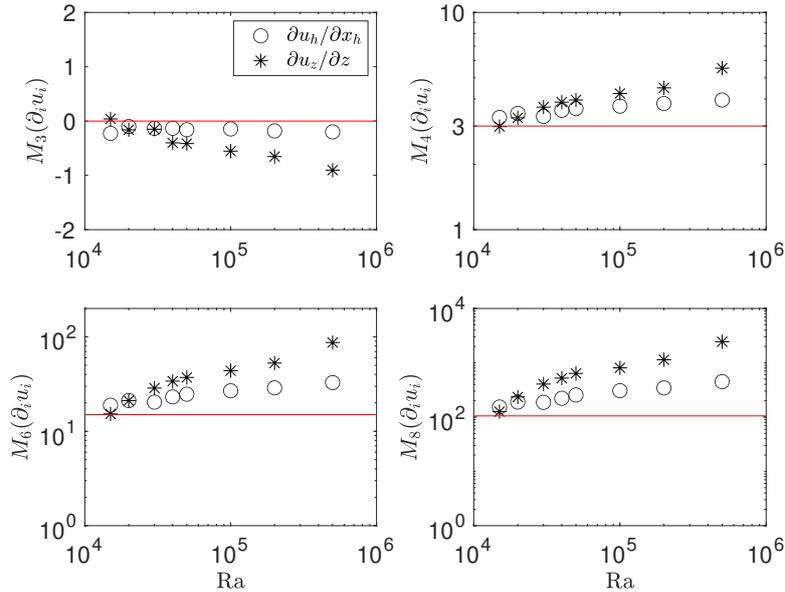}
\caption{Normalized derivative moments of different longitudinal derivatives versus Rayleigh number, see eq. \eqref{Mnf}. The Gaussian moment amplitudes of 0, 3, 15, and 105 for $n=3, 4, 6$ and 8, respectively, are indicated by solid lines in each panel. The horizontal longitudinal derivative $\partial u_h/\partial h$ is obtained as a combined moment of $\partial u_x/\partial x$ and $\partial u_y/\partial y$.  
\label{Mf}}
\end{center}
\end{figure}

The $n$th-order normalized derivative moments $M_n(\partial_i u_i)$, which are determined in the bulk (denoted to as $\langle\cdot\rangle_b$) of the convection layer between $2\delta_T$ and $1-2\delta_T$ (with the thermal boundary layer thickness $\delta_T=1/(2\Nu)$) and given by 
\begin{align}
M_n(\partial_i u_i)=\frac{\langle(\partial u_i/\partial x_i)^n\rangle_b}{\langle(\partial u_i/\partial x_i)^2\rangle_b^{n/2}}\,,
\label{Mnf}
\end{align}
are displayed in Fig. \ref{Mf} for orders $n=3,4,6$ and 8. If not stated otherwise, we omit the tilde for dimensionless quantities for the following. 

We plot moments of the longitudinal vertical derivative and the mean of the longitudinal horizontal derivatives which obey the same statistics. As a guide to the eye, we also indicate the corresponding magnitude which result for a Gaussian statistics, $M_n = n!!=1\cdot 3\cdot \dots (n-1)$ with $n$ even. All moments follow Gaussian statistics for $\Ra\sim 10^4$ and deviate increasingly to an intermittent statistics, particularly those of the component $\partial u_z/\partial z$. Such a phase transition from Gaussian to intermittent derivative statistics has been investigated in more detail in isotropic turbulence \cite{Yakhot2006,Schumacher2007,Biferale2008,Yakhot2017}, for Burgers turbulence \cite{Friedrich2018}, and for shear flows and thermal convection \cite{Schumacher2014,Schumacher2018,Yakhot2020}. While this transition is sharp in isotropic box turbulence \cite{Yakhot2017}, the crossover of derivative statistics from the Gaussian to the intermittent regime is gradual and seems to be altered by increasingly intermittently appearing coherent structures which grow out of the boundary layers (BLs) near the walls, such as the rising and falling thermal plumes in convection \cite{Schumacher2018,Yakhot2020,Valori}. Our goal here is to reveal the dynamical scenarios that are connected with the observed statistics as the larger magnitudes of the derivative moments of $\partial u_z/\partial z$ are connected to rising and falling thermal plumes.    
\begin{figure}
\begin{center}
\includegraphics[width=0.7\textwidth]{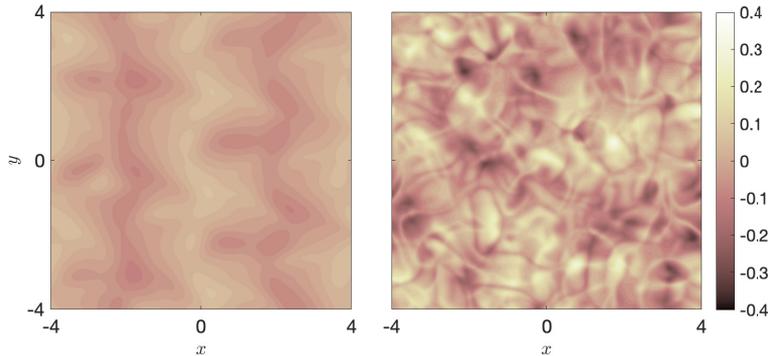}
\caption{Contrast field $\xi({\bm x}_2)$ which is defined by eq. \eqref{syn} for $\Ra=1.5\times 10^4$ (left) and  $5\times 10^5$ (right). In each case one example snapshot is taken. The colorbar is the same for both figures.   
\label{sync}}
\end{center}
\end{figure}
\begin{figure}
\begin{center}
\includegraphics[width=0.7\textwidth]{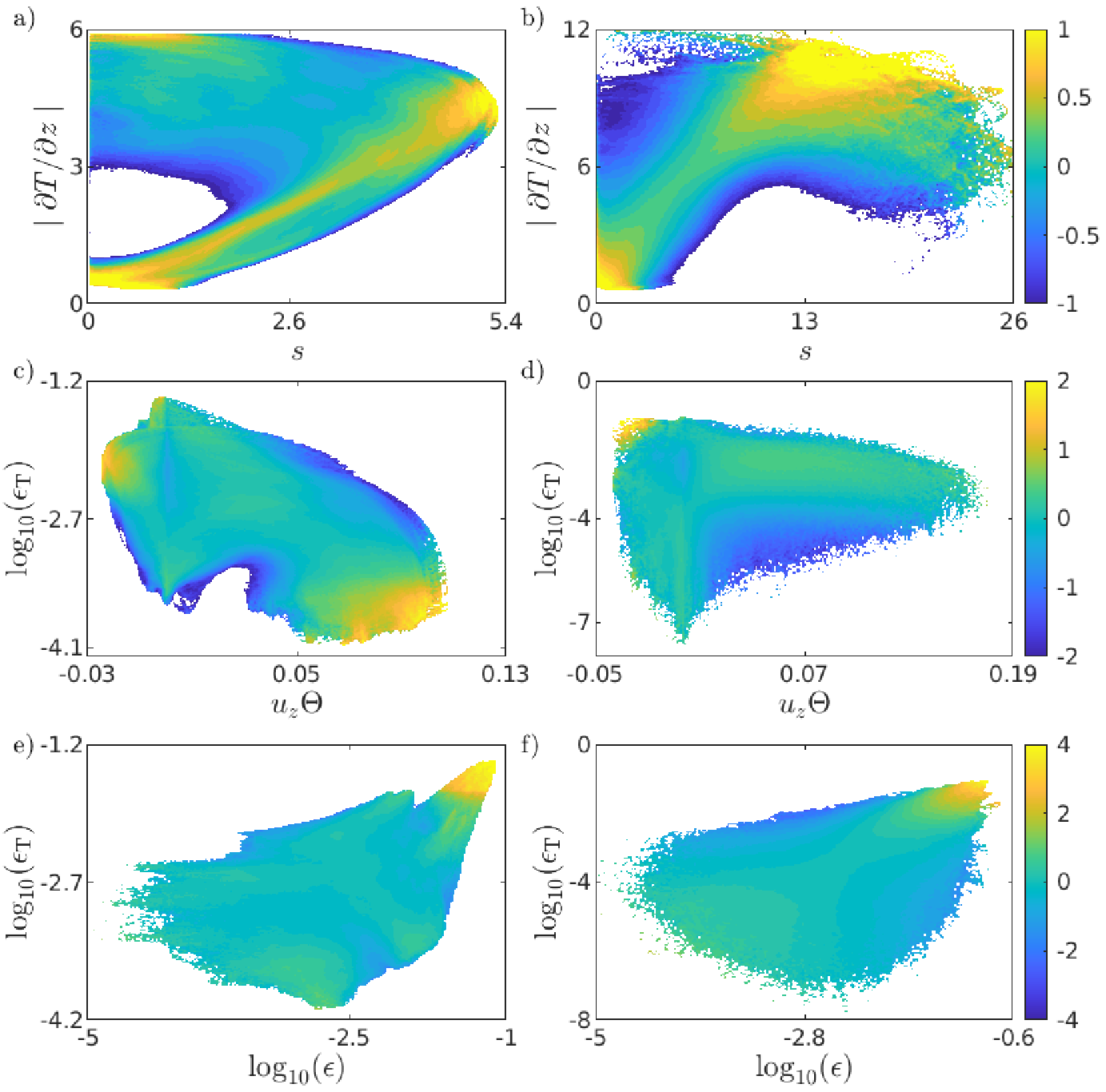}
\caption{Joint statistics at the walls and in the bulk of the convection layer, see eq. \eqref{syn}. Data in the left and right columns are for $\Ra=1.5\times 10^4$ (a,c,e) and $5\times 10^5$ (b,d,f), respectively. Panels (a,b) display the joint probability density function (JPDF) of the magnitudes of wall shear stress and vertical derivative of temperature at both walls. Panels (c,d) display the JPDF of the decadic logarithm of the thermal dissipation rate $\epsilon_T$ and the product $u_z\Theta$ in the midplane. Panels (e,f) connect the thermal dissipation rate with the kinetic energy dissipation rate $\epsilon$ in the midplane. The colorbar is the same for the two panels in each row, the contour level are given in decadic logarithm.   
\label{wall}}
\end{center}
\end{figure}

\section{Connection of boundary layers to bulk} In the weakly turbulent regime of convection, the convection rolls fill the whole layer. This implies that plume detachments at one plate are closely connected to plume impingements at the opposite one, both at nearly the same $(x,y)$. We term this point the {\em synchronicity} of the BL dynamics at the top and bottom. Figure \ref{sync} illustrates the behavior for two temperature snapshots at $\Ra=1.5\times 10^4$ (left) and  $5\times 10^5$ (right). We determine a contrast which can calculated  by (here ${\bm x}_2=(x,y)$)
\begin{align}
\xi({\bm x}_2)=[T({\bm x}_2,\delta)-\langle T\rangle] + [T({\bm x}_2,H-\delta)-\langle T\rangle]\,,
\label{syn}
\end{align}
where $\delta < \delta_T$ and $\langle T\rangle=1/2$. We see that the small Rayleigh number data are nearly perfectly synchronised while the ones for the largest $\Ra$ clearly underline the synchronicity breaking which allows for plume collisions in the bulk. This aspect is investigated in the following where we connect the boundary layer dynamics at the top and bottom to that in the bulk. Therefore, we will analyse data in three representative horizontal planes only, at $z=0, 1$  for the BLs and at $z=1/2$ for the bulk. This restriction is motivated by a planned future application of a similar analysis in laboratory experiments where data are obtained in very few light sheets only.   

The wall shear stress field is a two-dimensional vector field which is in physical dimensions given by 
\begin{equation}
{\bm \tau}_w=\rho_0 \nu \frac{\partial {\bm u}_t}{\partial n}\Bigg |_{\rm wall}\,,
\end{equation}   
that consists of the normal derivatives of the two tangential velocity components at the wall. Note that since the normal vector is pointed always inwards, the sign at the top and bottom walls will be opposite. In refs. \cite{Bandaru2015,Schumacher2016a,Pandey2018}, we discussed in detail the zero points, ${\bm \tau}_w=0$, and the different local topologies in their vicinity. It was shown there that saddle points are found at places where thermal plumes detach from the walls, unstable nodes are points where thermal plumes from the opposite walls impinge. These zero points form a network which is distributed across the plate. Thermal plumes detach in regions from the plates where the magnitude $|\partial T/\partial z|$  is small. 

Our subsequent statistical analysis is based on a sequence of snapshots that have been written out at a time interval $\Delta t =0.29$ for $\Ra =1.5\times 10^4$ and $\Delta t=0.475$ for $\Ra=5\times 10^5$. The first step to take is thus to study the joint statistics of wall shear stress magnitude and magnitude of the vertical temperature derivative. Rather than taking the wall shear stress, we follow the convention of \cite{Chong2012,Lenaers2012} and switch to the magnitude of the skin friction field $s=|{\bm s}|$ which is given in dimensional form by ${\bm s}={\bm \tau}_w/(\rho_0\nu)$.  In Figs. \ref{wall}(a,b) we display the collected statistics from both plates for two runs at $\Ra=1.5\times 10^4$ (left) and $5\times 10^5$ (right). In detail, we plot the joint normalized probability density function (JPDF) 
\begin{equation}
J(s, |\partial T/\partial z|)=\frac{p(s, |\partial T/\partial z|))}{p(s)\, p(|\partial T/\partial z|)}\,.
\end{equation}   
The additional normalisation highlights the strongly correlated values better. If $J>1$ then the correlation is larger than if the two fields were statistically independent. It is seen that for both Rayleigh numbers particularly high correlations follow (1) for the smallest skin friction field magnitudes in combination with plume detachments, where $|\partial T/\partial z|$ is very small (lower left corner of the data plane), and (2) for the largest skin friction field magnitudes with local maxima of $|\partial T/\partial z|$. The latter events are detected between saddle or node points \cite{Bandaru2015} (upper right corner of the data plane). The latter events seem to be statistically more significant, but our interest will be particularly on the events of type (1).   

\begin{figure}
\includegraphics[width=1\textwidth]{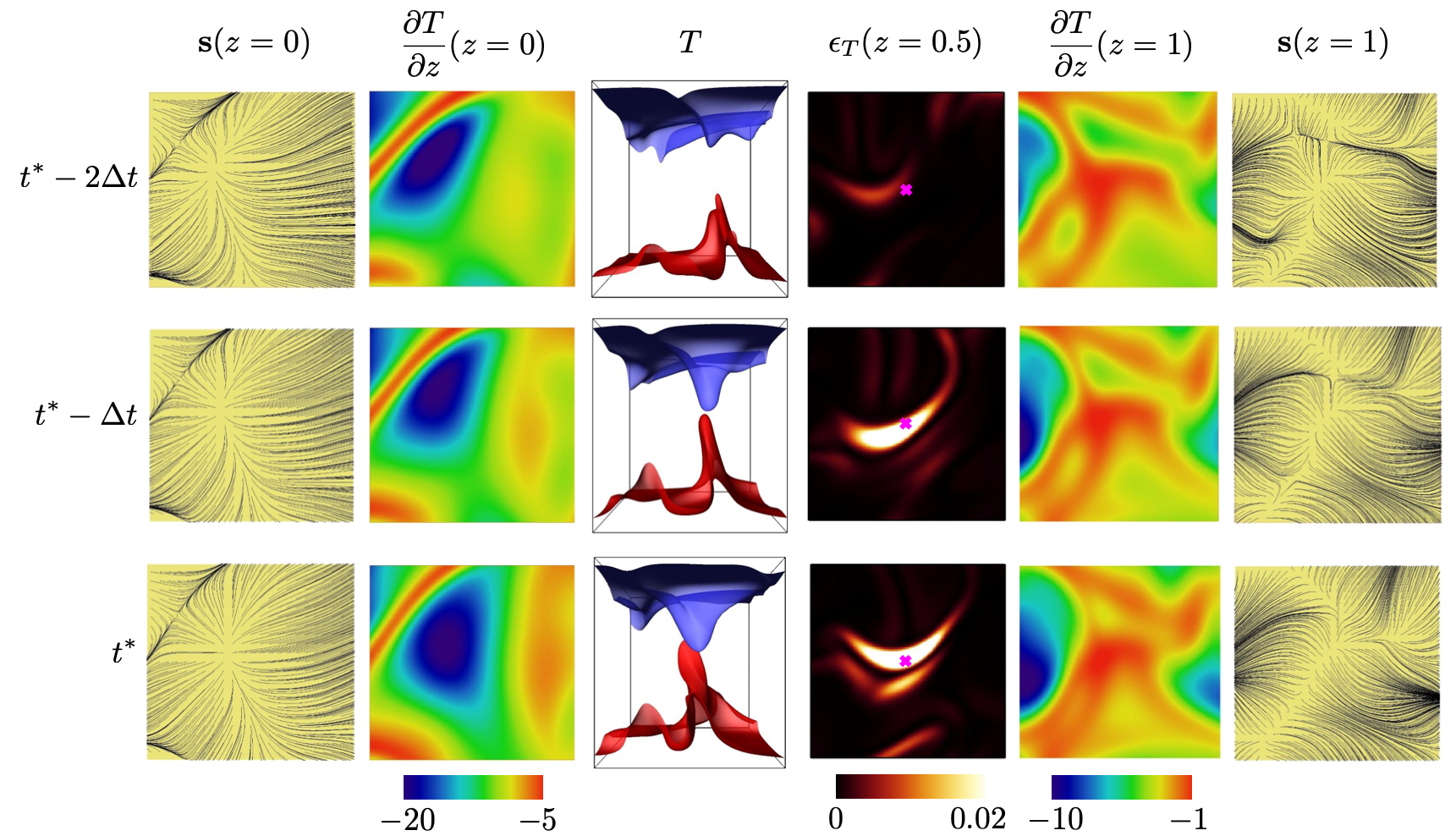}
\caption{Formation of an extreme thermal dissipation rate event in the centre of the convection layer. The panels in this figure illustrate the dynamical processes at the walls in connection with those in the bulk which lead to the event. The three rows from top to bottom stand for snapshots of relevant fields at times $t^\ast-2\Delta t$, $t^\ast-\Delta t$, and $t^\ast$ with $t^\ast$ the time at which the event occurs and $\Delta t=0.475$. The quantities which are displayed in the 6 columns are field lines of the skin friction field ${\bm s}$ at $z=0$, the vertical derivative of temperature, $\partial T/\partial z$ at $z=0$, the temperature $T$ across the layer, which is illustrated by two isosurfaces at $T=0.3$ (blue) and $T=0.7$ (red), the thermal dissipation rate $\epsilon_T$ at $z=0.5$, the vertical derivative of temperature at $z=1$, and field lines of the skin friction field at $z=1$. The area is always $H\times H$ and $Ra=5\times 10^5$. Color bars stand for all panels in a column.      
\label{genesis}}
\end{figure}
  
In Figs. \ref{wall}(c,d), we display $J(u_z\Theta, \epsilon_T)$ in the midplane that connects the local product of $u_z$ and $\Theta=T-\langle T\rangle_{V,t}$, a quantity that enters the convective heat current, and the thermal dissipation rate field which is given by 
\begin{equation}
\epsilon_T({\bm x},t)=\frac{1}{\sqrt{\Ra \Pra}}({\bm \nabla} T)^2\,. 
\label{epsT}
\end{equation}
We recall that the largest values $u_z\Theta>0$ correspond to rising or falling thermal plumes that reach the midplane \cite{Emran2012}. While for the lower Rayleigh number two regions with $J>1$ can be identified (one additional for the highest amplitude convective heat currents in combination with low-amplitude thermal dissipation rates) we detect for both Rayleigh numbers local maxima with negative $u_z\Theta$ and high thermal dissipation rate. Particularly for the highest $\Ra$, this maximum of $J$ is tight to the highest values of $\epsilon_T$. These are the events which we will be interested in, namely events that can be assigned to plume collisions or close passings that lead to reversed transport events, e.g., hot fluid moves from top to bottom. Such events are in line with the generation steep temperature gradients and thus with extreme bulk dissipation events. 

In Figs. \ref{wall}(e,f), connects finally the two dissipation rate fields of thermal dissipation rate (see eq. \eqref{epsT}) and kinetic energy dissipation rate. The latter field $\epsilon$ is given by  
\begin{equation}
\epsilon({\bm x},t)=\frac{1}{2}\sqrt{\frac{\Pra}{\Ra}}(\,{\bm \nabla} {\bm u}+({\bm \nabla} {\bm u})^T)^2\,.
\label{eps}
\end{equation}
The JPDFs for both Rayleigh numbers show a strong statistical correlation at their maximum amplitudes. This was for example shown in \cite{Scheel2013} for thermal convection in a closed cylindrical cell. The physical picture in connection with this correlation was discussed in detail in \cite{Schumacher2016}. Plume collisions in the bulk generate steep temperature gradients (and thus high amplitudes of $\epsilon_T$) and subsequently local shear layers (and thus high amplitudes of $\epsilon$). 

We can calculate the mean dissipation rates by the following exact relations which can be directly derived from the Boussinesq equations and which read for the chosen units and $\Pra=1$ as follows,
\begin{equation}
\langle\epsilon\rangle_{V,t}=\frac{\Nu -1}{\sqrt{\Ra}} \quad\mbox{and}\quad \langle\epsilon_T\rangle_{V,t}=\frac{\Nu}{\sqrt{\Ra}} \,.
\label{relations}
\end{equation}
The Nusselt numbers of $\Nu=2.52$ and 6.87 for $\Ra=1.5\times 10^4$ and $5\times 10^5$, respectively lead to the following mean dissipation rates: $\langle\epsilon\rangle_{V,t}=0.012$ and $0.0083$ for $\Ra=1.5\times 10^4$ and $5\times 10^5$, respectively. Furthermore,
$\langle\epsilon_T\rangle_{V,t}=0.021$ and $0.0097$ for $\Ra=1.5\times 10^4$ and $5\times 10^5$, respectively. The highest amplitudes, which we detect in the midplane, thus exceed the mean values of thermal dissipation by a factor of 18 for the lower Rayleigh number and a factor of 87 for the largest one in this series. In case of the kinetic energy dissipation rate, one obtains a factor of 14 and 23. Even though the Rayleigh numbers are still moderate, the distributions have developed already extended tails with high-amplitude events.         
       
Where do we stand in the present analysis? We have shown so far that the breaking of the synchronicity between the dynamics of the boundary layers at the top and bottom can cause plume detachments at nearly opposite positions $(x,y)$ and thus collisions in the bulk. This holds for the larger Rayleigh numbers only. Plume detachments at both walls are correlated to zero points of the skin friction field ${\bm s}$. Also for the higher of both $\Ra$ only, negative convective heat currents in the bulk, as they would be detected for plume collision events as the transport is topped and locally reversed, are connected to highest thermal dissipation amplitudes. The outlined dynamical scenario that leads to extreme bulk dissipation rate events is thus statistically confirmed by our analysis which concludes the next step in the present analysis.        

\section{Genesis of a concrete extreme event} In the final section of this work, we want to illustrate in detail the dynamics that leads to such an extreme (or high-amplitude) bulk dissipation event in the middle of the layer. It is one possible dynamical process that led to the highest dissipation rate amplitudes in our data record. In order to proceed, we use the same sequences of simulation data that were analysed in the past section statistically. We focus to an example event at the highest Rayleigh number and denote the time of the extreme bulk dissipation event by $t=t^{\ast}$.  

Figure \ref{genesis} illustrates the dynamics in a slab $H\times H\times H$ around the location of the extreme event of the thermal dissipation rate. Several quantities are displayed at output steps $t^{\ast}-2\Delta t$ to $t^{\ast}$. The figure covers roughly one free-fall time of the evolution of the turbulent convection flow. The outermost panels to the left and right show field lines of the skin friction vector field. At the bottom plate, we detect a plume impact region which is in line with a large magnitude of $\partial T/\partial z$ and an unstable node of ${\bm s}$. Furthermore, one observes a saddle point in the upper left corner that is connected to a low magnitude region of the vertical derivative of the temperature field. This is a plume detachment event. Similarly, one detects such processes at the top plate at $z=1$ in the outermost right column of Fig. \ref{genesis}. Saddle points or stable nodes are connected to low-magnitude regions of $\partial T/\partial z$. Also, there is an overlap of these regions at top and bottom when comparing the panels.  

Isosurface plots of the temperature field at 2 isolevels ($T=0.3$ and 0.7) illustrate the detachment of the plumes from the opposite walls. The simultaneous detachment does not necessarily lead to a frontal collision of both plumes as the panels show. Plumes that pass by each other closely will also generate strong local temperature gradients which are collected by the thermal dissipation rate. This dynamics is in line with an increase of $\epsilon_T$ in the midplane, as also shown in the figure (see the 4th column).      

\begin{figure}
\begin{center}
\includegraphics[width=0.8\textwidth]{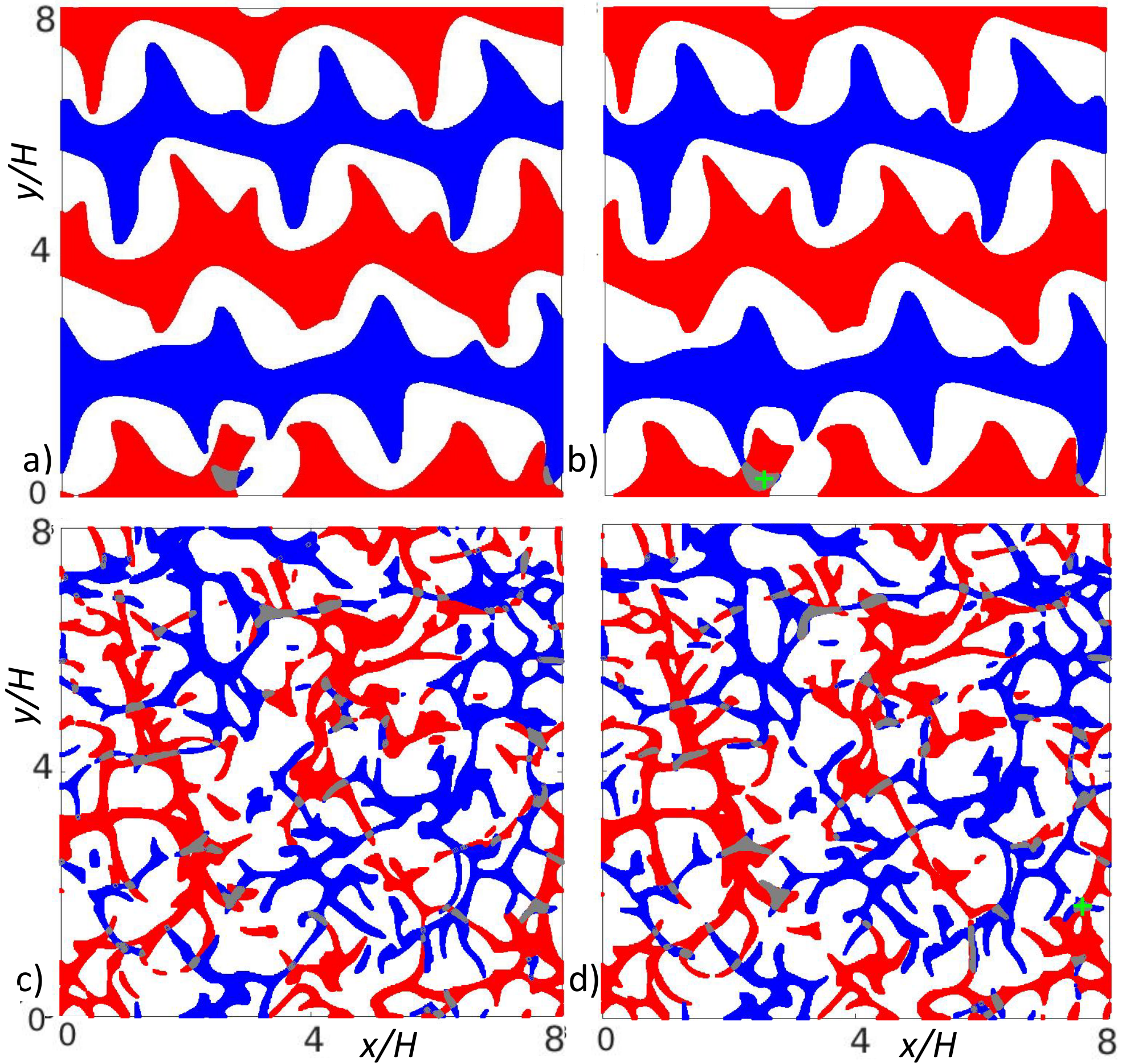}
\caption{Regions with $\partial T/\partial z > -A$ at the top (blue) and bottom (red) walls of the cell. Overlaps of the horizontal locations of $\partial T / \partial z > -A$ are coloured in grey. The horizontal position of the extreme event of $\epsilon_T$ at the mid-plane of the cell is indicated with a green cross. 
Top row: $ \Ra = 1.5\times 10^4$, $A = -1.5$, bottom row: $\Ra = 5\times 10^5$, $A = -2.5$. Left column: $t = t^*-\Delta t$, right column: $t = t^*$.}
\label{BLs-center}
\end{center}
\end{figure}

Figure \ref{BLs-center} shows the connection between $\partial T / \partial z$ at the top and bottom BLs and an extreme event of thermal energy dissipation rate in the bulk of the cell. It thus complements our previous analysis in Fig. \ref{sync}. Data are again reported for $\Ra$ number $1.5\times 10^4$, and $\Ra$ $5\times10^5$ in the top and bottom row, respectively. More specifically, this figure reports the position of the smallest values of $|\partial T / \partial z|$ on the whole plate, which are the preferred locations for the formation of thermal plumes from the boundary layers, at two time steps, $t^*-\Delta t$ in panels (a) and (c), and $t^*$ in panels (b) and (d). The horizontal position of the extreme event of thermal energy dissipation is indicated by a cross in panels b) and d). From the plots of the first row, one can observe again that at the lower $\Ra$ the regions where the plumes detach from the bottom boundary layer do not overlap with the regions of plume formation at the top one, except at the location of the extreme event. For the lower $\Ra$ number, the synchronicity between the top and the bottom of the cell is still nearly perfectly established, it starts to be broken only where the detected extreme event occurs. Such synchronicity is not visible for the largest $\Ra$ number in our series, where we can detect several intersections in the horizontal position of simultaneous plume detachments from the top and bottom boundary layers. The extreme event shown in panel (d) is the same event illustrated in Figure \ref{genesis}. The number of these events, which can be thought as precursors for extreme bulk dissipation events, grows steadily with increasing Rayleigh number and is the structural reason for the increasing deviations from a Gaussian derivative statistics in the bulk that we summarized in Fig. \ref{Mf}.  

\section{Summary and conclusion} We have studied the formation of extreme events of the derivatives of velocity and temperature in the bulk of a thermal convection layer which are probed by the rates at which kinetic energy and thermal variance are dissipated, $\epsilon$ and $\epsilon_T$. Our analysis is based on high-resolution direct numerical simulation data    
which are taken for $\Pra=1$ in a range of moderate Rayleigh numbers for which the derivative statistics in the bulk proceeds a crossover from Gaussian to intermittent behaviour. We demonstrated which dynamics in the boundary layers at the top and bottom can generate high dissipation events and thus intermittent statistics. It is shown that the breaking of the synchronicity  between the plume detachment dynamics at the top and bottom walls enhances the probability of plume collisions or close passings in the center of the convection layer and thus of dissipation rate events in the far tails of their distributions. The latter dynamical connection was demonstrated by a detailed analysis of the joint statistics and in a concrete example at the highest Rayleigh number. These intermittency generation mechanisms might change as we increase the Rayleigh number further. Their study has to be left as future work.    

Our analysis is a step for an experimental study of convective turbulence in gases at a Prandtl number $\Pra\sim 1$ which is currently prepared. In such a situation, the available measurement data is even more constrained as applied in the present case, where we restricted our analysis basically to three analysis planes on purpose. This holds particularly for the derivative statistics. On the one hand,  the statistics will lack some accuracy. On the other hand, not all 9 (3) components of the velocity gradient (temperature gradient) can be measured in space and time, except for three-dimensional Particle Image Velocimetry (PIV) or Particle Tracking Velocimetry (PTV) methods (combined PIV or PTV and optical temperature measurements).  A bridge to a partial solution of these challenges can however be built by the use of machine learning techniques that help to design models for the detection of precursors to extreme bulk dissipation events which use the present DNS data base as training data (see also ref. \cite{Pandey2020} for a discussion). Such studies are currently underway and will be reported elsewhere.  
   
\section{Acknowledgments}
The work is supported by the Priority Programme DFG-SPP 1881 on Turbulent Superstructures and the project  SCHU 1410/30 of the Deutsche Forschungsgemeinschaft. The authors gratefully acknowledge the Gauss Centre for Supercomputing e.V. (www.gauss-centre.eu) for funding this project by providing computing time through the John von Neumann Institute for Computing (NIC) on the GCS Supercomputer JUWELS at J\"ulich Supercomputing Centre (JSC).

\end{document}